\documentclass[sigconf,screen]{acmart}
\setcopyright{rightsretained}
\acmPrice{}
\acmDOI{10.1145/3611643.3616308}
\acmYear{2023}
\copyrightyear{2023}
\acmSubmissionID{fse23main-p591-p}
\acmISBN{979-8-4007-0327-0/23/12}
\acmConference[ESEC/FSE '23]{Proceedings of the 31st ACM Joint European Software Engineering Conference and Symposium on the Foundations of Software Engineering}{December 3--9, 2023}{San Francisco, CA, USA}
\acmBooktitle{Proceedings of the 31st ACM Joint European Software Engineering Conference and Symposium on the Foundations of Software Engineering (ESEC/FSE '23), December 3--9, 2023, San Francisco, CA, USA}
\received{2023-02-02}
\received[accepted]{2023-07-27}
\usepackage{amssymb,amsfonts}
\usepackage[T1]{fontenc}
\usepackage{algorithmic}
\usepackage{textcomp}
\usepackage{xargs}                      
\usepackage{url}
\usepackage{hyperref}
\usepackage{subcaption}
\usepackage[colorinlistoftodos,prependcaption]{todonotes}
\usepackage{dcolumn}
\usepackage[inline]{enumitem}                   
\usepackage[nameinlink]{cleveref}                   
\usepackage[nolist]{acronym}            
\usepackage{tcolorbox}                  
\usepackage{microtype}

\def\BibTeX{{\rm B\kern-.05em{\sc i\kern-.025em b}\kern-.08em
    T\kern-.1667em\lower.7ex\hbox{E}\kern-.125emX}}
\graphicspath{{figures/}}

\setlist[enumerate]{nosep}
\newcommandx{\mytodo}[2][1=]{\todo[inline,#1]{#2}}
\newcommand{\grey}{\color{gray}}

\DeclareMathOperator{\mab}{_{MAB}}

\newenvironment{takeaway}{
\begin{tcolorbox}\centering\itshape\vspace{-0.4\baselineskip}
}{
\vspace{-0.4\baselineskip}\end{tcolorbox}
}

\newcounter{point}
{}

\newcounter{Cpara}
\newcounter{Ipara}
\newcounter{Epara}

\renewcommand{\theCpara}{C\arabic{Cpara}}
\renewcommand{\theIpara}{I\arabic{Ipara}}
\renewcommand{\theEpara}{E\arabic{Epara}}

\newcommand\Cpar[1]{\par\refstepcounter{Cpara} \textbf{(\theCpara)\space#1.}}
\newcommand\Ipar[1]{\par\refstepcounter{Ipara} \textbf{(\theIpara)\space#1.}}
\newcommand\Epar[1]{\par\refstepcounter{Epara} \textbf{(\theEpara)\space#1.}}

\crefname{Cpara}{issue}{issues}
\crefname{Ipara}{issue}{issues}
\crefname{Epara}{issue}{issues}

\begin{document}
\title{Revisiting Neural Program Smoothing for Fuzzing}

\author{Maria-Irina Nicolae}
\email{Irina.Nicolae@bosch.com}
\affiliation{%
  \institution{Robert Bosch GmbH\\Bosch Center for AI}
  \city{Stuttgart}
  \country{Germany}
}
\author{Max Eisele}
\email{MaxCamillo.Eisele@bosch.com}
\affiliation{%
  \institution{Robert Bosch GmbH}
  \city{Stuttgart}
  \country{Germany}
}
\affiliation{%
\institution{Saarland University}
\city{Saarbr\"ucken}
\country{Germany}
}
\author{Andreas Zeller}
\email{zeller@cispa.de}
\affiliation{%
\institution{CISPA Helmholtz Center for Information Security}
\city{Saarbr\"ucken}
\country{Germany}
}


\begin{abstract}
    Testing with randomly generated inputs (fuzzing) has gained significant traction due to its capacity to expose program vulnerabilities automatically.
    Fuzz testing campaigns generate large amounts of data, making them ideal for the application of \ac{ML}.
    \emph{Neural program smoothing}, a specific family of \ac{ML}-guided fuzzers, aims to use a neural network as a smooth approximation of the program target for new test case generation.

    In this paper, we conduct the most extensive \emph{evaluation} of \ac{NPS} fuzzers against standard gray-box fuzzers (>11 CPU years and >5.5 GPU years), and make the following contributions:
    \begin{enumerate*} 
    \item We find that the original performance claims for \ac{NPS} fuzzers \emph{do not hold;} a gap we relate to fundamental, implementation, and experimental limitations of prior works.
    \item We contribute the first \emph{in-depth analysis} of the contribution of \acl{ML} and gradient-based mutations in \ac{NPS}.
    \item We implement Neuzz++, which shows that addressing the practical limitations of \ac{NPS} fuzzers improves performance, but
    that \emph{standard gray-box fuzzers almost always surpass \ac{NPS}-based fuzzers.}
    \item As a consequence, we propose \emph{new guidelines} targeted at benchmarking fuzzing based on \acl{ML}, and present MLFuzz, a platform with GPU access for easy and reproducible evaluation of \ac{ML}-based fuzzers.
    \end{enumerate*}
    Neuzz++, MLFuzz, and all our data are public.
\end{abstract}


\maketitle

\section{Introduction}
\label{sec:intro}


In recent years, fuzzing---testing programs with millions of random, automatically generated inputs---has become one of the preferred methods for finding bugs and vulnerabilities in software, mainly due to its speed, low setup efforts, and successful application in the industry.
Google's OSSFuzz initiative~\cite{google22OSSFuzz}, for instance, has revealed thousands of bugs in open-source software.

Fueled by success stories of practical fuzzing, researchers are constantly seeking ways to make fuzzers more efficient~\cite{li2018fuzzing}.
The most popular approach is still \emph{coverage-guided fuzzing:} generate new test cases from prior ones using an evolutionary search that optimizes code coverage through a fitness function.
Techniques used to enhance fuzzers include concolic execution~\cite{Yun2018, Stephens2016}, or static analysis~\cite{Wustholz2020}.
Along them, \acl{ML} methods have increasingly been applied to different parts of the fuzzing loop in academic research~\cite{godefroid2017,rajpal2017,chen2018,Drozd2018,bottinger2018}.

Fuzz testing generates significant amounts of data which make a welcome input for machine learning.
Moreover, obtaining labels through feedback from the fuzzer or the program is most often fast and cheap.
Constructing a dataset for training machine learning models is thus relatively straightforward in fuzzing.
However, despite their increased traction in the research community in the past decade, \ac{ML}-based fuzzers are not widely used in practice~\cite{Metzman2021}.

Recently, \emph{\acl{NPS}}~\cite{She2019,She2020,Wu2022} has been proposed to approximate the tested program with a neural network.
The trained model learns to predict coverage from test cases, being additionally smooth and differentiable.
These properties allow computing gradients, which cannot readily be done on programs directly.
Test cases are mutated into new ones based on the predictions of the neural network using gradient descent.
The use of gradients allows to steer the mutations in the most relevant directions, which have higher chances of reaching new coverage.
Despite promising significant performance gains, both in terms of code coverage and number of bugs found, these methods are not currently used by practitioners for testing real software.

Motivated by the applicability of \acl{NPS} to real-world fuzzing, we provide a \emph{systematic and thorough analysis of \ac{NPS}-guided fuzzing methods} with the following contributions:

\begin{enumerate} 
	\item We provide a critical analysis of \ac{NPS}-guided fuzzing, uncovering fundamental, conceptual and practical limitations that were previously ignored.
    We show that \emph{neural network performance does not translate to improved coverage,} as the model fails to capture rare edge coverage.
	\item We compare multiple \ac{NPS}-guided fuzzers in an extensive benchmark against AFL, AFL++, and the recent Havoc$\mab$ on 23 target programs.
    \ac{NPS}-guided fuzzers underperform regarding code coverage and bug finding, which is \emph{at odds with the results from the original papers.}
    We explain this performance gap by outdated or incorrect experimental practices in prior work.
    \item We reimplement Neuzz as a custom mutator for AFL++ and show that fixing practical limitations of \ac{NPS} significantly improves fuzzing performance.
    Nevertheless, we find that \acl{NPS} methods \emph{are outperformed by state-of-the-art gray-box fuzzers,} despite their use of additional computation resources.
    \item Based on our findings, we propose \emph{better-suited guidelines} for evaluating \ac{ML}-enhanced fuzzing, and present \emph{MLFuzz,} the first fuzzing benchmarking framework with GPU support dedicated to \ac{ML}-based fuzzing.
	\mbox{MLFuzz} allows for easy, reproducible evaluation of fuzzers with or without \acl{ML}, similar to standard practices used by FuzzBench~\cite{Metzman2021}.
\end{enumerate}

The remainder of the paper is structured as follows.
\Cref{sec:related} introduces prior work on coverage guided fuzzing and \acl{NPS}, before tackling our main analysis on limitations of \acl{NPS} in \Cref{sec:limitations}.
\Cref{sec:implem} presents our implementation of \ac{NPS} fuzzing and the benchmarking platform.
\Cref{sec:exp} covers experiments, followed by new experimental guidelines in \Cref{sec:guidelines}.
We conclude this work in \Cref{sec:conclusion}.
All our results and code are publicly available (\Cref{sec:data}).

\section{Background}
\label{sec:related}

\paragraph{Coverage-guided fuzzing}
Coverage-guided fuzzers explore the input space of a program starting from a few sample inputs called seeds.
They mutate the seeds into new test cases based on a \emph{fitness criterion,} which rewards reaching new code coverage obtained by gray-box access through binary instrumentation.
Test cases that increase coverage are kept in the corpus to be evolved further.
Over time, the input corpus and the total code coverage grow.
During execution, the fuzzer checks the target program for unwanted behavior, notably crashes and hangs.
Popular coverage-guided fuzzers are American Fuzzy Lop (AFL)~\cite{zalewski2017american}, its successor AFL++~\cite{Fioraldi2020}, and libFuzzer~\cite{libfuzzer}.
Alongside basic mutations, most gray-box fuzzers use the \emph{havoc} mutation strategy, where a fixed number of randomly chosen atomic mutations are chained to a more complex mutation~\cite{Fioraldi2020}.
Motivated by the success of havoc in modern fuzzers, Havoc$\mab$~\cite{Wu2022havoc} was designed to implement the havoc strategy as a two-layer multi-armed bandit~\cite{auer2002}.
Despite the trivial reward function used by the bandit, Havoc$\mab$ claims to significantly improve code coverage over random havoc in extensive benchmarks.

\paragraph{Fuzzing with \acl{ML}} 
\ac{ML} has been applied to various tasks in the fuzzing loop.
Neural byte sieve~\cite{rajpal2017} experiments with multiple types of recurrent neural networks that learn to predict optimal locations in the input files to perform mutations.
Angora~\cite{chen2018} uses byte-level taint tracking and gradient descent to mutate test cases towards new coverage.
FuzzerGym~\cite{Drozd2018} and B\"ottinger \emph{et al.}~\cite{bottinger2018} formulate fuzzing as a reinforcement learning problem that optimizes coverage.
In parallel to mutation generation, \acl{ML} is naturally fit for generating test cases directly.
Skyfire~\cite{wang2017} learns probabilistic grammars for seed generation.
Learn\&Fuzz~\cite{godefroid2017} uses a sequence-to-sequence model~\cite{Sutskever2014} to implicitly learn a grammar to produce new test cases.
GANFuzz~\cite{Hu2018} uses generative adversarial networks (GANs)~\cite{goodfellow2014} to do the same for protocols.
DeepFuzz~\cite{Liu2019} learns to generate valid C programs based on a sequence-to-sequence model for compiler fuzz testing.
The application of \ac{ML} to fuzzing is covered more extensively in~\cite{saavedra2019,wanga2019}.

\paragraph{Neural program smoothing}

\begin{figure}[t]
    \centering
    \includegraphics[width=\linewidth]{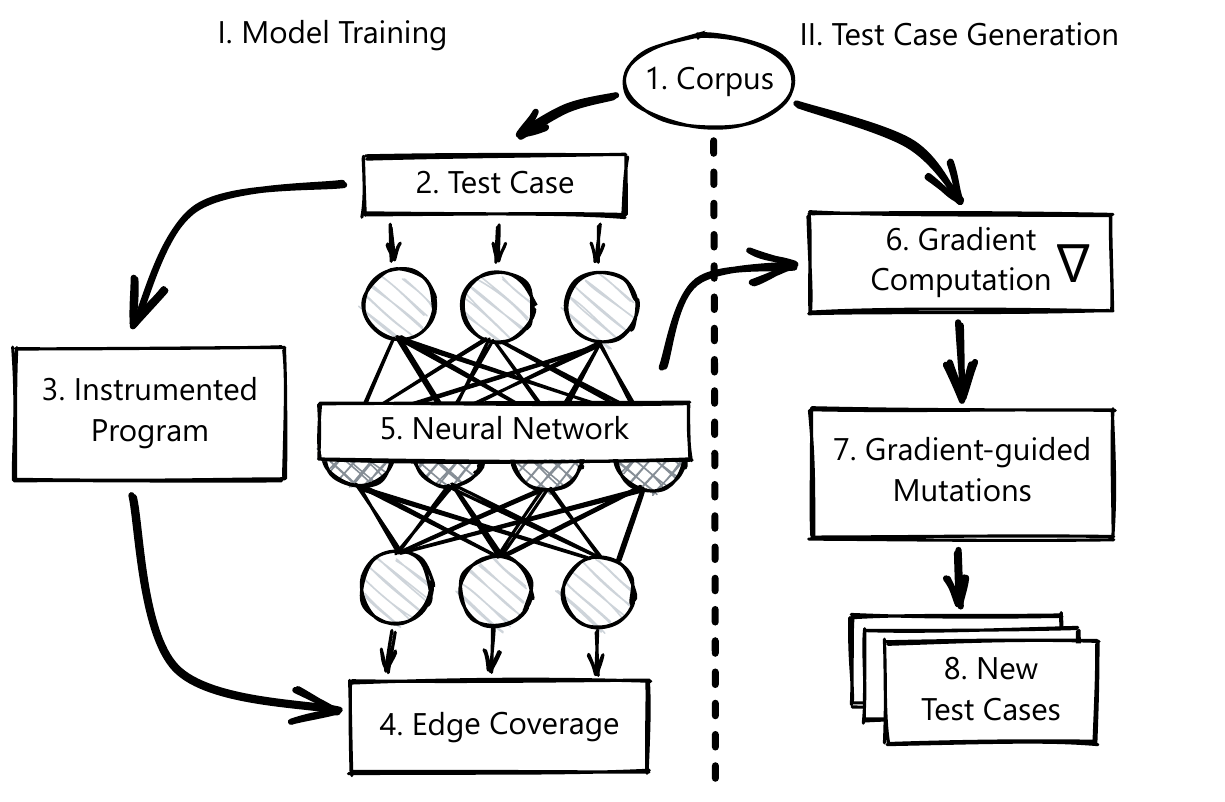}
    \caption{Neural program smoothing for fuzzing.}
    \label{fig:neural_program_smoothing}
\end{figure}

Program smoothing~\cite{Chaudhuri2010,Chaudhuri2011} was initially introduced as a way to facilitate program analysis and overcome the challenges introduced by program discontinuities.
Among the uses of \acl{ML} in fuzzing, neural program smoothing is one of the most recent and popular methods, due to its great performance in the original studies.
Neuzz~\cite{She2019} trains a neural network to serve as a smooth approximation of the original program in terms of code coverage (\Cref{fig:neural_program_smoothing}).
First, all test cases~(2) from the corpus~(1) are executed on the instrumented program~(3) to obtain their individual code coverage~(4), i.e. edge coverage from \texttt{afl-showmap}.
The respective pairs of test case and coverage are then used to train a neural network~(5), 
which learns to predict the coverage for each test case.
Being smooth and differentiable, the neural network can be used for computing \emph{gradients}, the values of derivatives of the program w.r.t.\ its inputs.
These indicate the direction and rate of fastest increase in the function value and can be used to flip specific edges in the bitmap from zero to one~(6).
Each gradient corresponds to one byte in the input.
The locations with the highest gradient values are mutated~(7) to propose new test cases~(8) that should reach the targeted regions of the code.
This idea is inspired by adversarial examples, more precisely FGSM~\cite{goodfellow2015}, where a change in the input in the direction of the sign of the gradient is sufficient to change the model outcome.

MTFuzz~\cite{She2020} extends Neuzz with multitask learning~\cite{caruana1997}: the neural network is trained against three types of code coverage instead of only edge coverage.
\emph{Context-sensitive coverage}~\cite{chen2018,wang2019} distinguishes between distinct caller locations for the same covered edge, while \emph{approach-sensitive coverage}~\cite{arcuri2010} introduces a third possible value in the coverage bitmap reflecting when an edge was nearly covered because the execution has reached a neighboring edge.
The three types of coverage help learn a joint embedding that is used to determine interesting bytes for mutation in the test case.
The bytes are ranked using a saliency score, which is computed as the sum of gradients for that byte in the learned embedding space.
Each ``hot byte'' is mutated by trying out all possible values, without further relying on the gradients.

PreFuzz~\cite{Wu2022} attempts to solve some limitations of Neuzz and MTFuzz by extending Neuzz in two ways.
The program instrumentation is changed to include all neighboring edges of covered ones in the bitmap.
This information is used to probabilistically choose which edge to target next for coverage, with the end goal of encouraging diversity in edge exploration.
Additionally, the success of havoc mutations~\cite{Fioraldi2020} is leveraged: after the standard Neuzz mutation, havoc is applied probabilistically to pre-defined segments of bytes in the test case, according to their gradient value.

\section{Analyzing Neural Program Smoothing}
\label{sec:limitations}

In this section, we provide our main analysis of \acl{NPS}, covering both the concepts behind \ac{NPS}, as well as existing fuzzer implementations.
We tackle three orthogonal perspectives: (i) conceptual or fundamental, (ii) implementation and usability, and (iii) experimental considerations.

\subsection{Conceptual Limitations}
\label{sec:limits_concept}

\Cpar{Approximation errors of the neural network} \label{lim:c1}
Being an empirical process, neural network training can suffer from errors introduced in the training process by, e.g., limited training data and training time, or sensitivity to hyperparameters.
Even in the ideal case, \emph{being a smooth approximation, the \ac{NPS} model will always differ from the actual program exactly at the most interesting points}, i.e.,\ discontinuities, branches, and jumps.
This approximation error is intrinsic to a smoothing approach and, at the same time, what allows \ac{NPS} methods to use gradients and numeric optimization towards producing new inputs.

\Cpar{Capacity to reach targeted edges} \label{lim:c2}
Arguably, the most salient research question to elucidate about \acl{NPS} is whether the gradient-guided mutation can indeed reach the targeted edges.
As \ac{NPS} is based on multiple components (\Cref{fig:neural_program_smoothing}), the overall performance of the fuzzer critically depends on the effectiveness of its individual components:
\begin{enumerate}
    \item The prediction accuracy of the neural network (5);
    \item The capacity of the gradient-based mutations (7) to achieve the expected new coverage on the target program.
\end{enumerate}

The experiments we perform later in the paper show that the \acl{ML} component as used by \acl{NPS} has impaired performance.
To the best of our knowledge, prior \ac{NPS} studies have not assessed what the model was learning and whether it was reaching its objective.

\Cpar{Incomplete coverage bitmaps} \label{lim:c3}
Another central limitation of \acl{NPS} that we uncover relates to the incompleteness of the coverage bitmaps that the neural network receives.
All \ac{NPS} fuzzers retrieve covered edges through \texttt{afl-showmap}, which only reports the edge IDs that are reached.
When the coverage information from all seeds is put together for the overall bitmap used for training the neural network, it thus only contains edges that were reached at least once by any of the seeds.
As such, unseen edges are not part of the bitmap and cannot be explicitly targeted and discovered by the model.
In practice, if the neural network does discover new edges, it is rather inadvertently due to randomness.
While having access to only an incomplete coverage bitmap is a conceptual limitation, it can be addressed on an implementation level.
It is sufficient to change the instrumentation of the program to include uncovered edges to overcome this issue.
Among existing \ac{NPS} fuzzers, PreFuzz is the only one that considers information about neighbors of reached edges in the coverage bitmap, albeit not motivated by the limitation we uncover.
Their goal is rather to be able to choose the next edge to target in a probabilistic fashion, depending on the degree of coverage of each edge and its neighbors.

The fundamental limitations uncovered in this section, while some easier to solve than others, are what we see as main obstacle in the adoption of \ac{NPS}-based fuzzing in practice.
As will be confirmed in \Cref{sec:exp}, the experiments are consistent with these limitations.

\subsection{Implementation and Usability Limitations}
\label{sec:limits_implem}

We now turn to practical aspects that make existing approaches to \acl{NPS} inconvenient to use, such that an independent evaluation requires major effort and code rewriting.

\Ipar{Use of outdated components} \label{lim:i1}
Existing implementations of \acl{NPS}~\cite{She2019,She2020,Wu2022}, along with Havoc$\mab$~\cite{Wu2022havoc} are implemented as extensions of AFL instead of using the more recent, more performant AFL++ as base.
Moreover, their dependency on outdated Python, TensorFlow and PyTorch versions impacts usability.
For the purpose of experiments, we have patched the code and updated the dependencies of all these fuzzers, as even for the most recent ones, some of their used libraries were already not available at the time of their publication.

\Ipar{Difficulty in building targets} \label{lim:i2}
Prior \ac{NPS} studies provided the binaries used in their own research, ensuring reproducibility.
However, for a fuzzer to be practical, it is advisable to rather provide instructions on how to build new programs for its use.
This is especially important when the fuzzer uses custom target instrumentation.
MTFuzz~\cite{She2020}, for instance, compiles a target program in five different ways due to the introduction of three additional types of instrumentation.
For this reason, we exclude MTFuzz from our empirical study as not being practical for real-world fuzzing.
Moreover, we argue that the three types of coverage used by MTFuzz are to a large extent redundant (conceptual limitation) and could be grouped into a unified coverage, thus reducing the build effort for this fuzzer.

\Ipar{Use of magic numbers} \label{lim:i3}
The magic numbers programming antipattern~\cite{martin2009} is frequently encountered in the implementations of \acl{NPS}-based fuzzers.
These values and other algorithmic changes are not mentioned in the original papers where each \ac{NPS} fuzzer is introduced.
It is thus difficult to establish whether the performance of each method is strictly linked to its proposed algorithm or rather to the implementation tweaks.
E.g., the maximum number of mutation guiding gradients per seed is set to 500; this value is not a parameter of the algorithm presented in the paper.

Our findings above show that the effort to set up existing \ac{NPS} fuzzers and build targets for them is significantly higher than for standard gray-box fuzzers, such as AFL and its variants, or libFuzzer.

\subsection{Evaluation Limitations}
\label{sec:limits_exp}

In this section, we highlight flaws and limitations of previous experimental evaluations of \ac{NPS} fuzzers and Havoc$\mab$, which have led to unrealistic performance claims.

\Epar{Experimental protocol} \label{lim:e1}
The more recent \ac{NPS} publications~\cite{She2020,Wu2022} \emph{lack of comparisons with recent gray-box fuzzers,} such as AFL++ and libFuzzer---fuzzers that were available and confirmed as state-of-the-art long before their publication.
Havoc$\mab$~\cite{Wu2022havoc} has included Neuzz and MTFuzz in their evaluation alongside AFL++.
However, we find that they use the same binary target for both AFL and AFL++, instead of building the program separately for AFL++.
AFL++ runs on AFL instrumented binaries, but not efficiently.
Moreover, the size of the coverage bitmap is usually larger for AFL++ than with AFL instrumentation; hence, code coverage as measured by the fuzzers is not directly comparable.
This makes the conclusions in the Havoc$\mab$ evaluation~\cite{Wu2022havoc} questionable.

\Epar{Fuzzer configuration for speed} \label{lim:e2}
We note that prior studies benchmarking \ac{NPS} methods compile their targets using \texttt{afl-gcc}, which results in slower targets and thus impacts fuzzing speed.
The AFL++ documentation recommends using preferably \texttt{afl-clang-fast} or \texttt{afl-clang-lto}~\cite{aflpp_guide}.
Additionally, AFL-based fuzzers have multiple options for transferring fuzz data to the program.
The most basic is to have AFL write test cases to file, and the target program executed with command line options to process the file as input.
The more sophisticated and recommended \emph{persistent} mode uses a fuzzing harness that repeatedly fetches fuzz data from AFL via shared memory and executes the function with the test data as input without restarting the whole program.
``\emph{All professional fuzzing uses this mode}'', according to the AFL++ manual~\cite{aydinbas2022aflpersistence}.
Depending on the target, the persistent mode can increase the throughput by 2--20$\times$~\cite{Fioraldi2020}.
Previous neural smoothing papers seem to run all experiments by feeding inputs via files, which should considerably slow down all fuzzers.
This is consistent with their results, where the more modern AFL++ consistently performs worse than AFL in the Havoc$\mab$ study~\cite{Wu2022havoc}, and the targets are printed with command line arguments in the original Neuzz paper~\cite{She2019}.
We conjecture that this tips the scale in favor of \ac{ML}-based fuzzers, which are themselves orders of magnitude slower than modern fuzzers~\cite{Drozd2018}.
This statement is validated experimentally in \Cref{sec:exp_slow}.

\section{Implementing Neuzz++ and MLFuzz}
\label{sec:implem}

In this section, we introduce \emph{Neuzz++,} our implementation of \acl{NPS} that aims to solve some limitations identified in \Cref{sec:limitations}, as well as the new experimental platform for evaluating \ac{ML}-based fuzzers.

\begin{figure}[t]
    \centering
    \includegraphics[width=\linewidth]{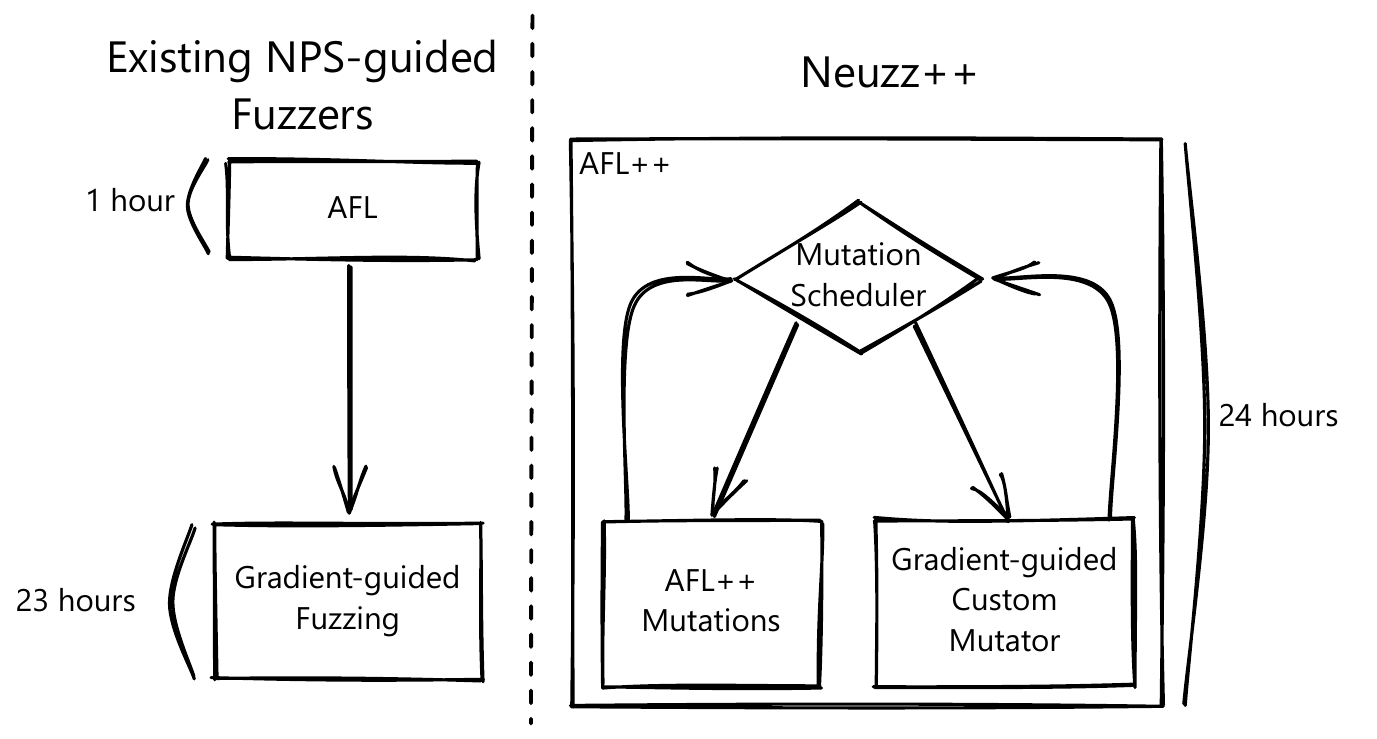}
    \caption{Operation mode of previous NPS-guided fuzzers and our Neuzz++.}
    \label{fig:neuzz++_implementation}
\end{figure}

\paragraph{Neuzz++}
We implement a variation of Neuzz as a custom mutator for AFL++, which we name Neuzz++ (see \Cref{fig:neuzz++_implementation}).
This allows our method to leverage most AFL++ features, like its standard mutations and power schedule.
More importantly, it allows for \acl{ML}-produced test cases and randomly mutated ones to evolve from each other.
We choose AFL++ as base for our implementation for its state-of-the-art performance, thus addressing \Cref{lim:i1}.
Being a custom mutator, Neuzz++ is modular, easy to build, and integrated with a default AFL++ installation.

In practice, Neuzz++ consists of two parts: the main AFL++ process with the custom mutator implemented in C, and a Python extension that is called for \acl{ML} operations.
The two processes communicate using named pipes.
We set a minimum requirement of $T$ test cases in the corpus for the custom mutator to run.
These are used to train the neural network for the first time; the model is retrained at most every hour if at least ten new test cases have been added to the corpus\footnote{Neuzz and PreFuzz solve this issue by running AFL for the first hour of fuzzing, then use the collected data for model training (\Cref{fig:neuzz++_implementation}).}.
This allows to refine the model over time with new coverage information from recent test cases.
In practice, we use $T=200$; this value is tuned experimentally and aims to strike the balance across all targets between fuzzing with \acl{ML} as early as possible, while waiting for enough data to be available for model training.
Intuitively, a larger dataset produces a better performing model.
\texttt{afl-showmap} is used to extract the coverage bitmap.
We introduce a coverage caching mechanism for model retraining which ensures that coverage is computed only for new test cases that were produced since last model training.
Each time the C custom mutator is called by AFL++, it waits for the Python component to compute and send the gradients of the test case.
Based on these, the mutations are computed by the C mutator and returned to AFL++.
In contrast to Neuzz, the gradients are not precomputed per test case, they are not saved to disk, the neural network is kept in memory, and the gradients are computed only on demand.
These optimizations minimize the time spent on \ac{ML}-related computations, keeping more time for fuzzing.

The neural network is a multi-layer perceptron (MLP) with the same structure as Neuzz (one hidden layer, 4096 neurons).
As shown in the PreFuzz paper~\cite{Wu2022}, we also found that different neural network architectures do not improve fuzzing performance.
In contrast to \ac{NPS} fuzzers, we keep 10\% of the test cases as validation set for evaluating the performance of the model.
We use the Adam optimizer~\cite{kingma2015}, a learning rate of $10^{-4}$, and cosine decay with restarts.

It is easy to parallelize model training and the main AFL++ routine for improved fuzzing effectiveness when testing real targets.
However, for experimental evaluation, we choose to have AFL++ wait for the neural network to train, similarly to previous implementations of neural program smoothing fuzzers.
This allows for fair experimental comparison and computation resource allocation.

The original Neuzz implementation applies four different mutation patterns on each byte selected according to the highest ranking gradients: incrementing the byte value until 255, decrementing the byte value down to 0, inserting a randomly sized chunk at the byte location, and deleting a randomly sized chunk starting at the given byte location.
We apply the same mutation pattern for Neuzz++. 

\paragraph{MLFuzz}
MLFuzz serves as a benchmarking framework for building test targets, running fuzzing trials in an isolated environment, and analyzing the findings.
Its main features are:

\begin{itemize}
	\item Test targets from Google Fuzzer Test Suite~\cite{Google2017} are compiled with the recommended and most recent compiler of the appropriate fuzzer; the build scripts are made available (addressing \Cref{lim:i2,lim:e2}).
	\item Targets are compiled with AddressSanitizer~\cite{serebryany2012} to detect memory errors.
    \item Six fuzzers are currently included in MLFuzz: AFL v2.57b, AFL++ v3.15a, Havoc$\mab$, Neuzz, PreFuzz and our Neuzz++.
    \item The implementation is containerized via Docker~\cite{merkel2014docker}. Python dependency specification is handled via virtual environments and Poetry~\cite{poetry2018}.
	\item Each fuzzing trial runs on one dedicated CPU and optionally one GPU for fuzzers that support it.
	\item All supported fuzzers have been modified to accept seeding of their random number generator for reproducible results.
	\item For all fuzzers, coverage is measured by replaying the corpus at the end of a run. We use binaries instrumented with AFL to ensure we do not disadvantage the AFL-based fuzzers, and \texttt{afl-showmap} from AFL++, since it has a larger bitmap with less hash collisions.
	\item Test cases are transmitted to fuzzers via shared memory, with the option to switch to slow transmission of test cases via the file system (addresses \Cref{lim:e2}).
\end{itemize}

\section{Experiments}
\label{sec:exp}

This section introduces our experiments and practical analysis, complementing the main findings from previous sections.
After presenting our setup (\Cref{sec:setup}), we assess the performance of the components of \ac{NPS}-based fuzzers in \Cref{sec:exp_ml_perf}.
We compare our Neuzz++ to prior \acl{NPS} fuzzers and standard gray-box fuzzers in an extensive benchmark in \Cref{sec:exp_perf}.
\Cref{sec:ml_cov,sec:qualitative,sec:exp_cpu} explore the added benefit of machine learning to \ac{NPS} fuzzers, while \Cref{sec:exp_slow} sheds light on experimental protocol differences with previous \ac{NPS} publications and their impact on fuzzing results.
Finally, we report bugs found in \Cref{sec:bugs}.

\subsection{Experimental Setup}
\label{sec:setup}

All experiments are performed on a server running Ubuntu 20.04 with four Nvidia Titan Xp GPUs.
Our study includes the six fuzzers from MLFuzz: AFL and AFL++ as standard gray-box fuzzers, Havoc$\mab$ as recent fuzzer claiming state-of-the-art performance, and \ac{NPS} fuzzers Neuzz, PreFuzz, and our own Neuzz++.
We use the original implementation and parameters provided by the authors for all baselines, except when stated otherwise.
We patch the code of Neuzz and PreFuzz to port them to Python 3.8.1, CUDA 11.5, TensorFlow 2.9.1~\cite{tensorflow2015} and PyTorch 1.4~\cite{pytorch2019}, as the original implementations are based on outdated libraries that are not available anymore or incompatible with our hardware.

\begin{table}
    \caption{Target programs from Google Fuzzer Test Suite~\cite{Google2017} and FuzzBench~\cite{Metzman2021}.}
    \label{tab:targets}
    \small
    \centering
    \begin{small}
    \begin{tabular}{lcrr}
    \toprule
    Target & Format & Seeds$^{\mathrm{a}}$ & LOC$^{\mathrm{b}}$ \\
    \midrule
    \multicolumn{4}{l}{\textbf{Source: Fuzzer Test Suite}} \\
    \midrule
    boringssl-2016-02-12 & SSL private key & 107 & 102793 \\
    freetype2-2017 & TTF, OTF, WOFF & 2 & 95576c \\
    guetzli-2017-3-30 & JPEG & 2 & 6045 \\
    harfbuzz-1.3.2 & TTF, OTF, TTC & 58 & 21413 \\
    json-2017-02-12 & JSON & 1 & 23328 \\
    lcms-2017-03-21 & ICC profile & 1 & 33920 \\
    libarchive-2017-01-04 & archive formats & 1 &  141563 \\
    libjpeg-turbo-07-2017 & JPEG & 1 & 35922 \\
    libpng-1.2.56 & PNG & 1 & 24621 \\
    libxml2-v2.9.2 & XML & 0 & 203166 \\
    openssl-1.0.2d & DER certificate & 0 & 262547 \\
    pcre2-10.00 & PERL regex & 0 & 67333 \\
    proj4-2017-08-14 & custom & 44 & 6156 \\
    re2-2014-12-09 & custom & 0 & 21398 \\
    sqlite-2016-11-14 & custom & 0 & 122271 \\
    vorbis-2017-12-11 & OGG & 1 & 17584 \\
    woff2-2016-05-06 & WOFF & 62 & 2948 \\
    \midrule
    \multicolumn{4}{l}{\textbf{Source: FuzzBench}} \\
    \midrule
    bloaty & ELF, Mach-O, etc. & 94 & 690642 \\
    curl & comms. formats & 41 & 153882 \\
    libpcap & PCAP & 1287 & 56663 \\
    openh264 & H.264 & 174 &  97352 \\
    stb & image formats & 467 & 71707 \\
    zlib & zlib compressed & 1 & 30860 \\
    \bottomrule
    \multicolumn{4}{l}{$^{\mathrm{a}}$Targets that do not have seeds use the default from Fuzzbench.} \\
    \multicolumn{4}{l}{$^{\mathrm{b}}$Retrieved with \texttt{cloc}~\cite{Danial2021}.} \\
    \end{tabular}
    \end{small}
\end{table}

We choose Google Fuzzer Test Suite~\cite{Google2017} and FuzzBench~\cite{Metzman2021} as standard, extensive benchmarks for our experimental evaluation.
We make use of 23 targets, summarized in \Cref{tab:targets}.
These are selected for being accessible, having dependencies available on Ubuntu 20.04, and being non-trivial to cover through fuzz testing.
Note that we only include targets from FuzzBench if they are not already included in Fuzzer Test Suite.
All results are reported for 24 hours of fuzzing.
We repeat each experiment 30 times to account for randomness, unless stated otherwise.
Each standard gray-box fuzzer is bound to one CPU core, while \ac{NPS} fuzzers are allotted one CPU and one GPU per trial. 
The main metrics used for evaluation are code coverage and number of bugs found.
For code coverage, we use edge coverage as defined by the AFL family of fuzzers.
However, we emphasize that AFL and AFL++ compute edge coverage differently.
In order to avoid the measuring errors introduced when ignoring this aspect, we count coverage by replaying the corpus using \texttt{afl-showmap} from AFL++ on the same binary, independently of which fuzzer was used in the experiment.
The setup we use fixes all experimental limitations we highlighted in \Cref{sec:limits_exp} (\Cref{lim:e1,lim:e2}).

\subsection{Performance of Machine Learning Models}
\label{sec:exp_ml_perf}

\begin{table}
    \caption{Dataset properties and neural network evaluation.}
    \label{tab:nn_perf}
    \centering
    \begin{small}
    \begin{tabular}{lrrrrrr}
    \toprule
    Target & \%covered edges & Acc & Prec & Recall & F1 & PR-AUC \\
    \midrule
    bloaty & 17.1\% & 0.53 & 0.17 & 0.18 & 0.17 & 0.15 \\
    boringssl & 19.3\% & 0.90 & 0.18 & 0.17 & 0.17 & 0.20 \\
    curl & 15.2\% & 0.89 & 0.15 & 0.15 & 0.15 & 0.23 \\
    freetype2 & 8.6\% & 0.89 & 0.09 & 0.09 & 0.09 & 0.10 \\
    guetzli & 18.5\% & 0.84 & 0.18 & 0.18 & 0.18 & 0.19 \\
    harfbuzz & 6.9\% & 0.93 & 0.07 & 0.07 & 0.07 & 0.07 \\
    json & 12.7\% & 0.88 & 0.11 & 0.08 & 0.09 & 0.10 \\
    lcms & 20.9\% & 0.84 & 0.19 & 0.19 & 0.19 & 0.21 \\
    libarchive & 6.9\% & 0.94 & 0.07 & 0.06 & 0.06 & 0.07 \\
    libjpeg & 17.8\% & 0.84 & 0.17 & 0.09 & 0.17 & 0.18 \\
    libpcap & 6.4\% & 0.92 & 0.06 & 0.06 & 0.06 & 0.07 \\
    libpng & 28.8\% & 0.86 & 0.28 & 0.27 & 0.27 & 0.29 \\
    libxml2 & 10.5\% & 0.92 & 0.10 & 0.09 & 0.09 & 0.11 \\
    openh264 & 21.4\% & 0.81 & 0.22 & 0.30 & 0.21 & 0.22 \\
    openssl & 31.2\% & 0.79 & 0.30 & 0.30 & 0.29 & 0.31 \\
    pcre2 & 4.3\% & 0.96 & 0.04 & 0.03 & 0.03 & 0.04 \\
    proj4 & 8.2\% & 0.95 & 0.08 & 0.07 & 0.07 & 0.08 \\
    re2 & 16.2\% & 0.87 & 0.15 & 0.13 & 0.13 & 0.16 \\
    sqlite & 16.3\% & 0.91 & 0.12 & 0.12 & 0.12 & 0.17 \\
    stb & 6.0\% & 0.92 & 0.06 & 0.05 & 0.05 & 0.06 \\
    vorbis & 29.6\% & 0.81 & 0.30 & 0.30 & 0.30 & 0.30 \\
    woff2 & 22.8\% & 0.85 & 0.22 & 0.22 & 0.21 & 0.13 \\
    zlib & 16.1\% & 0.85 & 0.14 & 0.10 & 0.11 & 0.16 \\
    \bottomrule
    \end{tabular}
    \end{small}
\end{table}

\begin{figure*}[t]
    \centering
    \includegraphics[width=\linewidth]{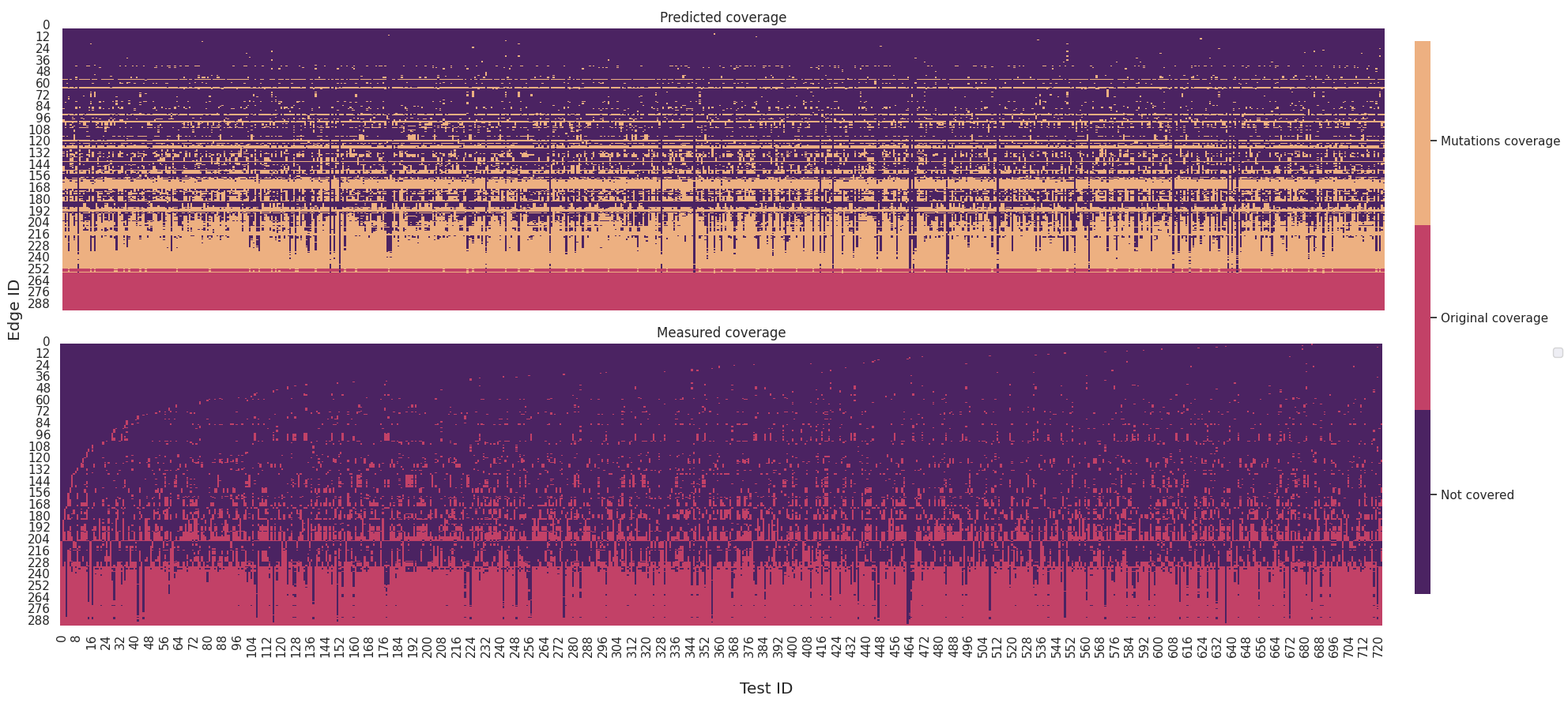}
    \vspace{-0.5\baselineskip}
    \caption{Predicted and actual edge coverage on \emph{libpng} for the entire corpus. Top: \ac{ML}-predicted coverage (pink) is trivial and almost constant over test cases. When each edge is targeted by mutations, predicted coverage (orange) increases for certain edges, but many code edges remain unattainable. Bottom: Coverage extracted with \texttt{afl-showmap} shows that all edges present have been covered at least once by the corpus.}
    \label{fig:libpng_cov}
\end{figure*}

We now investigate the quality of coverage predictions by the neural network and gradient-based mutations, in relation to concerns about the fundamental principle of \acl{NPS} (\Cref{sec:limits_concept}).
We tackle the following questions:
\begin{itemize}
    \item Can the neural network learn to predict edge coverage?
    \item Can gradient-based mutations reach targeted edges?
\end{itemize}
To this end, we propose quantitative and qualitative analyses of the performance of the neural network in \acl{NPS} fuzzers.
Without loss of generality, we investigate these based on Neuzz++ as a proxy for all \acl{NPS} fuzzers included in our study.
As all these methods use the same neural network architecture, loss function, method of training, etc., it is to be expected that their models will achieve the same performance when trained on the same dataset.
The results of the analyses can be summarized as follows and are detailed subsequently:
\begin{itemize}
    \item \Cref{tab:nn_perf} quantifies the model performance for all targets in terms of standard \acl{ML} metrics;
    \item \Cref{fig:libpng_cov} provides a qualitative analysis of model predictions for a given target, opposing them to correct labels.
    \item Lastly, \Cref{fig:libpng_cov} also assesses the capacity of the neural network to reach edges through gradient-based mutations.
\end{itemize}

\paragraph{\ac{ML} performance metrics}
To assess one factor of difficulty of the \acl{ML} task, we evaluate dataset imbalance for the training corpus.
This measures the percentage of the positive class (covered edges, in our case the minority) in the coverage bitmap of the training set.
Recall that the bitmap is produced by \texttt{afl-showmap} and accounts for the coverage obtained by the corpus before training; the coverage was not necessarily achieved based on a neural network, but rather by AFL++ mutations.
Note that this value is averaged across test cases and edges; rare edges might have much smaller coverage ratios, resulting in more difficulty in training an accurate model for those edges.
When facing class imbalance, the model tends to prefer the majority class, thus making wrong predictions.
For this reason, the performance of the neural network is assessed using precision, recall, F1-score, and \ac{PR} trade-off as performance metrics for the neural network.
Accuracy is also computed for completeness, but keep in mind that this metric is misleading for imbalanced datasets\footnote{One can trivially predict all-zeros (no coverage) and obtain very high accuracy.}.
We measure the \ac{AUC} of the \ac{PR} metric to evaluate all the operational points of the neural network.
Similar to accuracy, \ac{PR}-\ac{AUC} saturates at one, but is more sensitive to wrong predictions in the positive class.
The learning setup of \acl{NPS} is a multi-label binary classification task, i.e., for each test case, multiple binary predictions are made, one per edge; in consequence, the metrics are computed for each edge in the bitmap independently, then averaged over all edges, and finally averaged over trial repetitions.

\Cref{tab:nn_perf} reports the model performance metrics, along with the percentage of the positive class in the dataset as imbalance metric.
All model metrics are computed on a 10\% holdout set of test cases that were not used for training.
As Neuzz++ retrains the model multiple times, all measurements are performed on the last trained neural network using the state of the corpus at that time.
The precision, recall, F1-score, and \ac{PR}-\ac{AUC} values in \Cref{tab:nn_perf} indicate that the neural network has low performance.
These metrics are particularly low when the class imbalance is stronger, i.e., for small values of ``\%covered edges''.
The dataset imbalance is quite extreme for seven targets, where the positive class represents less than 10\% of the dataset, making predictions particularly difficult.

To provide an intuition into what the neural network learns, we design a qualitative evaluation of its predicted coverage.
This experiment uses the target \emph{libpng} and the test cases generated in a 24-hours run of Neuzz++.
\Cref{fig:libpng_cov} shows two coverage plots for this target for the entire corpus, where each ``column'' in the plot represents one test case, while each ``row'' is a program edge.
We compare the coverage predicted by a trained \ac{ML} model for the same test cases and edges (\Cref{fig:libpng_cov} top) to the true coverage extracted with \texttt{afl-showmap} (bottom).
The bottom plot is the coverage bitmap extracted with \texttt{afl-showmap} for the corpus and used for model training by Neuzz, PreFuzz, and Neuzz++.
A reduction (deduplication) operation is applied to it, which for \emph{libpng} reduces the number of edges from ~900 to the 293 present in the plot; this operation also explains any visual artifacts present in the image, as the edges are reordered.
The pink areas of the two plots differ significantly, with the model predictions being almost constant over all test cases: the model only predicts trivial coverage and fails to capture rare edges.
While this is a consequence of the difficulty of the \acl{ML} tasks (small dataset, class imbalance, too few samples w.r.t.\ the size of the test cases and bitmaps, see \Cref{tab:nn_perf}), it results in large approximation errors in the neural network, as outlined in \Cref{lim:c1}.
Moreover, recall that Neuzz, PreFuzz and Neuzz++ use the same \ac{ML} model type and structure, with minor differences in the training procedure and similar model performance.
Our findings thus extend to all \ac{NPS} methods.

Finally, we investigate the effectiveness of gradient-based mutations as essential component of \ac{NPS} fuzzers.
In the same setup on \emph{libpng} from the previous section, we apply Neuzz++ mutations to the corpus generated by a 24-hours fuzzing run as follows.
For each edge in the bitmap, we consider the case when it is explicitly targeted and generate all mutations with a maximum number of iterations in the mutation strategy.
\Cref{fig:libpng_cov} (top) plots the predicted coverage for each test case and edge before the mutations, as well as the increment of coverage after mutation.
Each edge (row) is considered covered by one test case (column) if at least one of the few thousand mutations generated to target it reaches the code location.
The results represent coverage estimated by the \ac{ML} model, not run on the program.
However, the coverage the model predicts is an optimistic estimate of the one actually achieved on the target, as the model dictated the mutations.
Note that the mutations are generated in the same way for Neuzz, PreFuzz and Neuzz++; our analysis thus applies to all methods and targets.

\Cref{fig:libpng_cov} (top) indicates that some locations are more readily reachable through mutations.
The harder to reach edges overall match the rarer edges of the corpus, as measured by \texttt{afl-showmap} in the bottom plot.
Most importantly, \emph{none of the edges targeted or covered by the mutations in the top plot represent new coverage}.
Recall that, by \ac{NPS} methods' design, a code edge is only present in the bitmap only if it has already been covered by the initial corpus used for training (\Cref{lim:c3}).
This becomes evident in the bottom plot of \Cref{fig:libpng_cov}: all edges have been covered by at least one test case.
As will be shown later, this fundamental flaw of \ac{NPS} methods translates to a limited practical capacity of reaching new coverage.

\begin{takeaway}
    The model predicts trivial edge coverage (\Cref{lim:c1}), and gradient mutations cannot target new edges (\Cref{lim:c3}).
\end{takeaway}

\subsection{Comparing Code Coverage}
\label{sec:exp_perf}

\begin{figure}
    \centering
    \includegraphics[width=\linewidth]{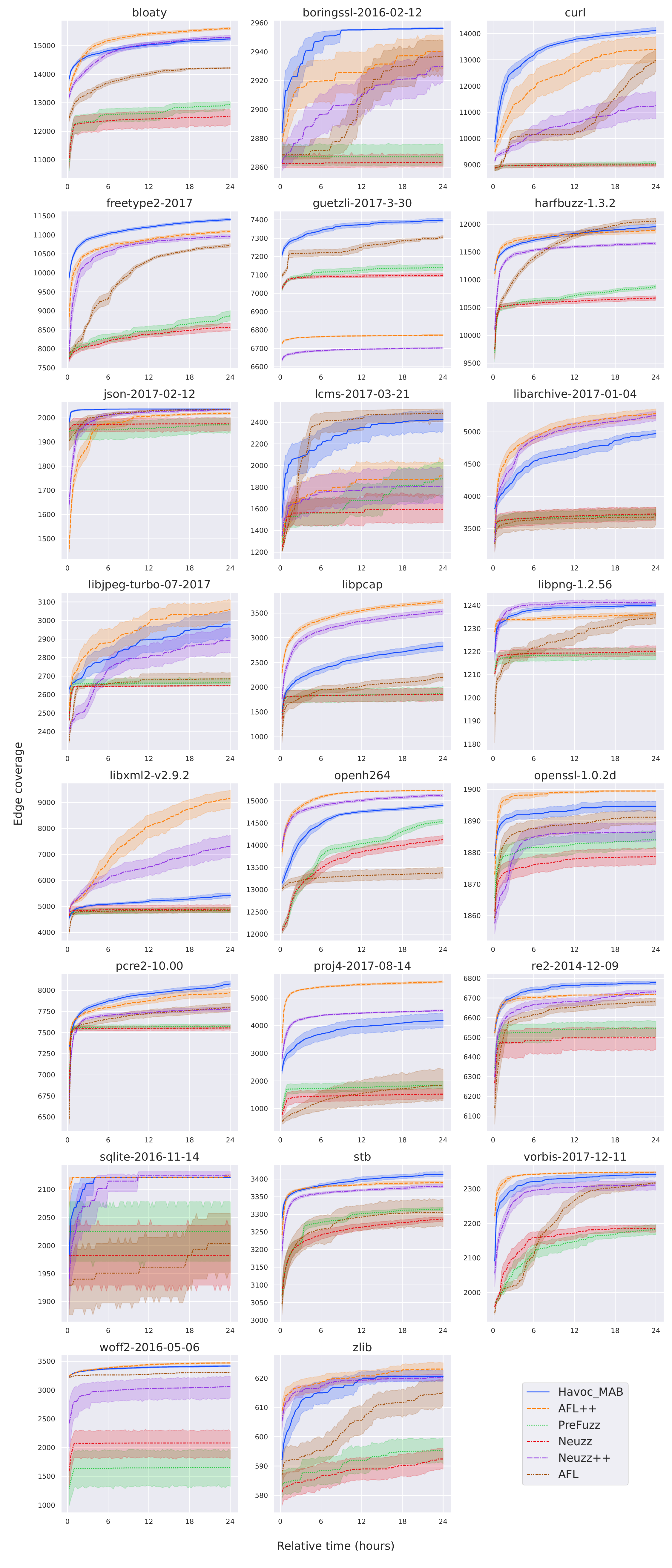}
    \caption{Average edge coverage over time with 95\% confidence interval.}
    \label{fig:cov}
\end{figure}

We present the main experiment comparing the achieved code coverage of available neural program smoothing approaches to AFL, AFL++ and the recent Havoc$\mab$ in \Cref{tab:cov} (average coverage) and~\Cref{fig:cov} (coverage over time).
This experiment alone requires a total computation time of over 11 CPU years and 5.5 GPU years.

\begin{table*}
\begin{minipage}[t]{.69\textwidth}%
\small
\centering
\caption{Average edge coverage and standard deviation over 30 runs.\\Best value in bold, second best underlined.}
\label{tab:cov}
\begin{tabular}{l D{:}{{\color{gray}\pm}}{5.3} D{:}{{\color{gray}\pm}}{5.3} D{:}{{\color{gray}\pm}}{5.3} D{:}{{\color{gray}\pm}}{5.3} D{:}{{\color{gray}\pm}}{5.3} D{:}{{\color{gray}\pm}}{5.3}}
\toprule
\multicolumn{1}{c}{Target} & \multicolumn{1}{c}{AFL} & \multicolumn{1}{c}{AFL++} & \multicolumn{1}{c}{Havoc$\mab$} & \multicolumn{1}{c}{Neuzz} & \multicolumn{1}{c}{PreFuzz} & \multicolumn{1}{c}{\textbf{Neuzz++}} \\
\midrule
bloaty & 14220:\grey49 & \textbf{15607}:\grey100 & 15240:\grey194 & 12518:\grey790 & 12936:\grey319 & \underline{15296}:\grey196 \\
boringssl & 2936:\grey34 & \underline{2940}:\grey34 & \textbf{2956}:\grey1 & 2863:\grey14 & 2867:\grey21 & 2930:\grey32 \\
curl & 13002:\grey1103 & \underline{13398}:\grey1376 & \textbf{14121}:\grey324 & 8999:\grey211 & 9048:\grey218 & 11260:\grey1401 \\
freetype2 & 10722:\grey126 & \underline{11090}:\grey104 & \textbf{11408}:\grey95 & 8569:\grey281 & 8870:\grey386 & 10960:\grey138 \\
guetzli & \underline{7306}:\grey21 & 6772:\grey9 & \textbf{7398}:\grey29 & 7099:\grey26 & 7141:\grey45 & 6702:\grey7 \\
harfbuzz & \textbf{12056}:\grey137 & 11887:\grey137 & \underline{11953}:\grey146 & 10672:\grey110 & 10875:\grey102 & 11654:\grey65 \\
json & \underline{2033}:\grey5 & 2018:\grey11 & \textbf{2036}:\grey0 & 1974:\grey79 & 1970:\grey86 & 2032:\grey12 \\
lcms & \textbf{2483}:\grey198 & 1904:\grey441 & \underline{2423}:\grey277 & 1593:\grey372 & 1876:\grey433 & 1809:\grey455 \\
libarchive & 3708:\grey383 & \textbf{5281}:\grey207 & 4970:\grey153 & 3729:\grey289 & 3718:\grey225 & \underline{5246}:\grey204 \\
libjpeg & 2685:\grey82 & \textbf{3058}:\grey161 & \underline{2980}:\grey192 & 2647:\grey5 & 2664:\grey60 & 2892:\grey189 \\
libpcap & 2203:\grey219 & 3733:\grey115 & 2833:\grey243 & 1859:\grey358 & 1875:\grey373 & 3529:\grey155 \\
libpng & 1234:\grey7 & 1235:\grey3 & \underline{1240}:\grey2 & 1220:\grey6 & 1219:\grey7 & \textbf{1241}:\grey3 \\
libxml2 & 4857:\grey286 & \textbf{9155}:\grey989 & 5416:\grey273 & 4895:\grey403 & 4853:\grey271 & \underline{7306}:\grey1191 \\
openh264 & 13381:\grey341 & \textbf{15234}:\grey15 & 14902:\grey109 & 14135:\grey241 & 14537:\grey142 & \underline{15126}:\grey80 \\
openssl & 1891:\grey6 & \textbf{1899}:\grey1 & \underline{1894}:\grey4 & 1878:\grey7 & 1884:\grey8 & 1886:\grey8 \\
pcre2 & 7797:\grey142 & \underline{7960}:\grey142 & \textbf{8076}:\grey98 & 7555:\grey76 & 7575:\grey77 & 7763:\grey67 \\
proj4 & 1837:\grey1621 & \textbf{5585}:\grey101 & 4190:\grey741 & 1526:\grey514 & 1849:\grey392 & \underline{4550}:\grey78 \\
re2 & 6680:\grey52 & 6717:\grey7 & \textbf{6777}:\grey26 & 6497:\grey162 & 6547:\grey110 & \underline{6731}:\grey30 \\
sqlite & 2004:\grey154 & \underline{2123}:\grey12 & 2121:\grey0 & 1982:\grey158 & 2025:\grey146 & \textbf{2125}:\grey16 \\
stb & 3305:\grey109 & \underline{3390}:\grey11 & \textbf{3413}:\grey20 & 3286:\grey18 & 3315:\grey14 & 3380:\grey14 \\
vorbis & 2317:\grey6 & \textbf{2348}:\grey4 & \underline{2342}:\grey19 & 2186:\grey29 & 2181:\grey40 & 2311:\grey38 \\
woff2 & 3305:\grey3 & \textbf{3472}:\grey34 & \underline{3418}:\grey36 & 2080:\grey667 & 1650:\grey893 & 3062:\grey530 \\
zlib & 615:\grey12 & \textbf{623}:\grey6 & \underline{620}:\grey5 & 592:\grey10 & 595:\grey12 & \underline{620}:\grey3 \\
\bottomrule
\end{tabular}
\end{minipage}
\hfill
\begin{minipage}[t]{.3\textwidth}%
\small
\centering
\caption{Average edge coverage of ML component over 30 runs.}
\label{tab:cov_ml}
\begin{tabular}{D{:}{{\color{gray}\pm}}{3.3} D{:}{{\color{gray}\pm}}{3.3} D{:}{{\color{gray}\pm}}{3.3}}
\toprule
\multicolumn{1}{c}{Neuzz} & \multicolumn{1}{c}{PreFuzz} & \multicolumn{1}{c}{\textbf{Neuzz++}} \\
\midrule
292:\grey206 & \textbf{666}:\grey593 & 261:\grey166 \\
0:\grey2 & 0:\grey0 & \textbf{11}:\grey19 \\
141:\grey182 & 153:\grey88 & 534:\grey245 \\
635:\grey225 & \textbf{865}:\grey255 & 429:\grey130 \\
28:\grey11 & \textbf{69}:\grey35 & 32:\grey13 \\
161:\grey52 & \textbf{377}:\grey145 & 345:\grey163 \\
2:\grey3 & 21:\grey46 & \textbf{216}:\grey86 \\
61:\grey210 & \textbf{344}:\grey415 & 8:\grey23 \\
105:\grey97 & 101:\grey124 & \textbf{1136}:\grey206  \\
4:\grey5 & 6:\grey6 & \textbf{103}:\grey35 \\
43:\grey27 & 59:\grey41 & \textbf{1608}:\grey289\\
2:\grey3 & 1:\grey2 & \textbf{105}:\grey31 \\
52:\grey46 & 74:\grey32 & \textbf{950}:\grey215 \\
1846:\grey290& \textbf{2248}:\grey284 & 148:\grey67 \\
6:\grey7 & 6:\grey6 & \textbf{91}:\grey70 \\
0:\grey2 & 13:\grey9 & \textbf{1396}:\grey160\\
186:\grey242 & 145:\grey204 & \textbf{197}:\grey82 \\
25:\grey93 & 24:\grey75 & \textbf{660}:\grey209 \\
0:\grey0 & 0:\grey0 & \textbf{94}:\grey129 \\
122:\grey77 & \textbf{150}:\grey82 & 77:\grey13 \\
\textbf{186}:\grey21 & \textbf{186}:\grey42 & 16:\grey11 \\
5:\grey16 & 14:\grey49 & \textbf{39}:\grey21 \\
9:\grey9 & 10:\grey8 & \textbf{24}:\grey14 \\
\bottomrule
\end{tabular}
\end{minipage}
\end{table*}

Overall, AFL++ obtains the best performance on ten targets, followed by Havoc$\mab$ with eight targets, and Neuzz++ on par with AFL, winning two targets each.
In view of AFL++ performance w.r.t.\ AFL, it is clear that not including AFL++ as a baseline in all prior \acl{NPS} works leads to overly optimistic conclusions about their capacities.
After AFL++, Havoc$\mab$ is the second most performant fuzzer in terms of code coverage.
However, we find that it does not reach the expected ranking advertised in the Havoc$\mab$ paper~\cite{Wu2022havoc}.

We observe that Neuzz and PreFuzz are never in the top two fuzzers.
Moreover, although they were designed to improve AFL performance, their coverage is in most cases lower than that of AFL.
AFL wins on 20 out of 23 targets over Neuzz, and 18 out of 23 over PreFuzz.
PreFuzz outperforms Neuzz on most targets, however this difference is significant only on six targets (see confidence intervals in \Cref{fig:cov}).
This finding is also at odds with original PreFuzz results~\cite{Wu2022}, where the performance gap is significantly wider.
\Cref{sec:exp_slow} is dedicated to further explaining the difference in performance with the initial papers.
Neuzz++ obtains higher coverage than Neuzz and PreFuzz on 21 programs, proving that our improvements over these methods are effective.

Targets \emph{libarchive, libxml2, proj4,} and \emph{woff2} exhibit the most variability among fuzzers.
Neuzz and PreFuzz exhibit large standard deviation on \emph{woff2,} where coverage varies depending if the fuzzers reach plateau or not.
For the other targets, it seems AFL-based fuzzers do not perform as well as AFL++-based ones.

\begin{takeaway}
    Overall, AFL++ achieves the highest code coverage. Among \ac{NPS} fuzzers, Neuzz++ achieves the highest code coverage.
\end{takeaway}

\subsection{Code Coverage from Machine Learning}
\label{sec:ml_cov}

After presenting total coverage for 24-hour runs in \Cref{tab:cov}, we now measure how much of the total coverage can be attributed to the machine learning component for each \ac{NPS} fuzzer.
On one hand, the goal is to discount the coverage produced strictly by AFL in the first hour of Neuzz and PreFuzz runs (recall that they use AFL for data collection, see \Cref{fig:neuzz++_implementation}) and only measure the \ac{NPS} fuzzers' contribution.
On the other hand, we wish to do the same for Neuzz++, and separate its contribution from that of the base fuzzer AFL++.
As Neuzz++ is a custom mutator for AFL++, its seeds usually alternate with regular AFL++ seeds.
To this end, we measure edge coverage by corpus replaying, this time only taking into account the seeds obtained by Neuzz, PreFuzz and Neuzz++, respectively.
For Neuzz and PreFuzz, this is equivalent to excluding the first hour of coverage, as done by the original authors.
In practice, this will include \ac{ML}-based mutations, but also other hard-coded mutations that the methods apply, such as havoc in the case of PreFuzz.
\Cref{tab:cov_ml} summarizes the comparison of edge coverage obtained by the \ac{ML} components of Neuzz, PreFuzz, and Neuzz++.
Program names are aligned with \Cref{tab:cov}.

Neuzz++ obtains the highest coverage in 14 over 23 targets, with values at least one order of magnitude higher than Neuzz and PreFuzz.
Nevertheless, even on targets where Neuzz++ does not obtain the highest \ac{ML} coverage (e.g., \emph{freetype2, harfbuzz}), the overall Neuzz++ edge coverage (\Cref{tab:cov}) is higher than that of Neuzz and PreFuzz, with the latter two obtaining lower coverage than their base fuzzer AFL.
The added value of Neuzz and PreFuzz is low in nine, respectively five targets, with coverage close to zero.
In these cases, Neuzz and PreFuzz do not achieve (almost) any coverage past the first hour of fuzzing with AFL (see also \Cref{fig:cov}).
This reinforces our previous conclusion that the time spent using Neuzz and PreFuzz might be better spent applying the AFL or AFL++ mutation strategy.
Moreover, the Neuzz++ results suggest that it might benefit from the alternation between \ac{ML}-guided mutations and standard AFL++ ones.
We explore this last point with additional analyses in \Cref{sec:qualitative}.

\begin{takeaway}
   For most programs, the time budget spent on Neuzz or PreFuzz is better spent on standard gray-box fuzzing.
\end{takeaway}

\subsection{Quality of Machine Learning Test Cases}
\label{sec:qualitative}

\begin{table}
    \caption{Statistics for \ac{ML}-generated test cases of Neuzz++. ``\%ML seeds'' and ``\%derived'' are computed over the total size of the corpus. ``\%MLcov+'' is relative to ``\%ML seeds''.}
    \label{tab:ml_seeds}
    \small
    \centering
    \begin{tabular}{lrrr}
    \toprule
    Target & \%ML seeds & \%MLcov+ & \%derived \\
    \midrule
    bloaty & 4.72\% & 28.3\% & 8.78\% \\
    boringssl & 27.7\% & 3.3\% & 27.5\% \\
    curl & 18.6\% & 26.1\% & 33.1\% \\
    freetype2 & 2.2\% & 31.9\% & 3.8\% \\
    guetzli & 9.9\% & 9.8\% & 13.8\% \\
    harfbuzz & 6.6\% & 30.2\% & 15.2\% \\
    json & 13.7\% & 37.3\% & 25.4\% \\
    lcms & 1.6\% & 57.1\% & 1.1\% \\
    libarchive & 18.3\% & 30.2\% & 34.9\% \\
    libjpeg & 11.8\% & 10.7\% & 15.9\% \\
    libpcap & 13.8\% & 40.0\% & 20.8\% \\
    libpng & 19.6\% & 13.8\% & 41.0\% \\
    libxml2 & 15.1\% & 23.9\% & 30.1\% \\
    openh264 & 10.2\% & 8.0\% & 8.5\% \\
    openssl & 30.6\% & 5.6\% & 28.7\% \\
    pcre2 & 18.3\% & 17.4\% & 29.88\% \\
    proj4 & 5.5\% & 49.7\% & 7.4\% \\
    re2 & 23.3\% & 22.8\% & 34.7\% \\
    sqlite & 8.1\% & 20.2\% & 6.2\% \\
    stb & 14.8\% & 15.3\% & 19.9\% \\
    vorbis & 6.0\% & 8.3\% & 7.9\% \\
    woff2 & 3.8\% & 19.8\% & 5.2\% \\
    zlib & 16.9\% & 20.7\% & 18.1\% \\
    \bottomrule
    \end{tabular}
\end{table}

We now aim to assess the quality of the test cases found by the \acl{ML} component of Neuzz++.
We do so with two analyses: we investigate (i) the inclusion of \ac{ML}-generated inputs in the AFL++ power schedule for further mutation, and (ii) the rarity of code edges found through \acl{ML}-based mutations.

First, \Cref{tab:ml_seeds} presents statistics regarding \ac{ML}-produced test cases for each target averaged over all trials.
The column ``\%ML seeds'' shows the overall percentage of inputs produced through \ac{ML} mutations.
Out of these, ``\%MLcov+'' discover new coverage (relative percentage).
Finally, ``\%derived'' is the total percentage of the corpus produced by direct mutations of \ac{ML}-based inputs.
We find that the ratio of machine learning inputs varies significantly across targets, representing up to a third of the corpus.
\ac{ML} test cases seem to be most impactful for finding new coverage on programs where they represent a low percentage of the corpus.
On average, each \ac{ML} test case is mutated at least once successfully, generating new test cases that are kept by Neuzz++ in the corpus.

The second analysis studies whether \ac{NPS} fuzzers explore code areas that are harder to reach by standard fuzzers.
In that case, \acl{NPS} fuzzers could be used in an ensemble of diverse fuzzers, opening the path for all fuzzers to rare parts of the code~\cite{Chen2019}.
To measure the rarity of edges reached by Neuzz++, we compare the edge IDs that Neuzz++ and AFL++ reach on each program, all trials joint.
The edge IDs are obtained by replaying all the test cases with \texttt{afl-showmap}.

\begin{table}[t]
    \caption{Reached edges for AFL++ (A+) and Neuzz++ (N+).}
    \label{tab:edge_intersection}
    \centering
    \begin{small}
    \begin{tabular}{l@{    }rrrrrrrrrr}
    \toprule
    \textbf{Target} & A+ & N+ & A+ $\cup$ N+ & A+ $\setminus$ N+ & N+ $\setminus$ A+ \\ 
    \midrule
    bloaty & 5131 & 4926 & 5139 & 213 & 8 \\ 
    boringssl & 1210 &	1208 &	1210 &	2 &	0 \\ 
    curl & 6862 & 6656 & 6899 & 243 & 37 \\ 
    freetype2 & 6555 &	6292 &	6575 &	283 \\ 
    guetzli & 2645 &	2624 &	2655 &	31 &	10 \\ 
    harfbuzz & 5438 &	5081 &	5440 &	359 &	2 \\ 
    json & 2036 &	2036 &	2036 &	0 &	0 \\ 
    lcms & 935 & 	1010 &	1019 &	9 	& 84 \\ 
    libarchive  & 3622 &	3223 &	3649 &	426 &	27 \\ 
    libjpeg-turbo & 1565 &	1539 &	1565 &	26 & 	0 \\ 
    libpcap & 2869 & 2628 & 2901 & 273 & 32 \\ 
    libpng & 621 &	616 &	621 &	5 &	0 \\ 
    libxml2 & 5401 &	4856 &	5410 &	554 &	9 \\ 
    openh264 & 5849 & 5840 & 5851 & 11 & 2 \\ 
    openssl & 812 &	811 &	812 &	1 &	0 \\ 
    pcre2 & 5548 & 5355 & 5577 & 222 & 29 \\ 
    proj4 & 2802 &	2343 &	2803 &	460 &	1 \\ 
    re2 & 2391 &	2426 &	2440 &	14 &	49 \\ 
    sqlite & 950 &	950 &	950 &	0 &	0 \\ 
    stb & 2012 & 2014 & 2021 & 7 & 9 \\ 
    vorbis & 1142 &	1100 &	1142 &	42 &	0 \\ 
    woff2  & 1254 &	1268 &	1270 &	2 &	16 \\ 
    zlib & 337 & 333 & 337 & 4 & 0 \\ 
    \bottomrule
    \end{tabular}
    \end{small}
\end{table}

We summarize the results in \Cref{tab:edge_intersection} as follows: Neuzz++ (denoted N+) reveals less than 0.5\% additional edges that AFL++ (denoted A+) in 16 out of 23 targets.
Neuzz++ does not find any such exclusive edges for eight programs; it is most successful on \emph{lcms,} with 8.2\% exclusive edges. 
On the other hand, AFL++ finds up to 16.4\% exclusive edges, lacking exclusive edges on only two programs (\emph{json} and \emph{sqlite}).
We can therefore conclude that \ac{NPS}-guided fuzzers explore essentially the same code areas as traditional fuzzers.

\begin{takeaway}
    \ac{NPS} fuzzers find less rare edges than gray-box fuzzers.
\end{takeaway}

\subsection{\ac{NPS}-based Fuzzing without GPUs}
\label{sec:exp_cpu}

\begin{table}
    \caption{Average edge coverage of \ac{NPS} fuzzers with and without GPU access (10 runs).}
    \label{tab:cpu_gpu}
    \centering
    \begin{small}
    \begin{tabular}{lrrrrrr}
    \toprule
    & \multicolumn{2}{c}{Neuzz} & \multicolumn{2}{c}{PreFuzz} & \multicolumn{2}{c}{\textbf{Neuzz++}} \\
    Target & CPU & GPU & CPU & GPU & CPU & GPU \\
    \midrule
    harfbuzz & 10607 & \textbf{10677} & 10675 & \textbf{10897} & 11631 & \textbf{11664} \\
    libjpeg & 2618 & \textbf{2647} & 2635 & \textbf{2688} & \textbf{2998} & 2892 \\
    sqlite & \textbf{2017} & 1993 & 2049 & \textbf{2089} & 2121 & \textbf{2127} \\
    woff2 & \textbf{1919} & 1816 & \textbf{1702} & 954 & 3288 & \textbf{3389} \\
    \bottomrule
    \end{tabular}
    \end{small}
\end{table}

Due to their increased performance for linear algebra and data throughput, GPUs are the \emph{de facto} standard for machine learning.
All \ac{NPS} methods studied in this paper leverage GPU access to train \acl{ML} models and compute gradients for test case mutations.
In practice, this means that they use more computational resources than state-of-the-art gray-box fuzzers, and that practitioners are required to invest in additional hardware.
In this section, we wish to assess the performance of \ac{NPS} methods in the absence of GPUs.
Model training with only CPU access should be slower, but it should not impact the performance of the trained model.
As such, any loss in fuzzing performance comes from spending more time training and less fuzzing.
For this small experiment, we select four targets that operate on a varied range of input formats for diversity.
We perform ten trials of all \ac{NPS} fuzzers with and without GPU access (\Cref{tab:cpu_gpu}).

Nine of twelve experiments obtain more code coverage when training the model on GPU, which is to be expected.
The exception is PreFuzz on woff2, which is however aligned with this fuzzer's tendency of sometimes becoming stuck on this program (\Cref{tab:cov}).
Overall, the fuzzing performance on GPU is marginally better, as training times for \ac{NPS} models are relatively short.
The gap between CPU and GPU seems tighter for Neuzz++, which we attribute to an already optimized and short training procedure, which cannot be much further improved by GPUs.

\begin{takeaway}
    Using GPUs usually results in better coverage for \ac{NPS}~fuzzers.
\end{takeaway}

\subsection{Impact of Test Case Transmission Method}
\label{sec:exp_slow}
In \Cref{lim:e2}, we underlined that \ac{NPS}-guided fuzzers use files to transfer test cases to the target program.
We now show that test case transmission has a major impact on fuzzing performance for the methods in~\cite{She2019, She2020, Wu2022havoc, Wu2022}.
We note that AFL++ does not reach the performance of its predecessor AFL by a margin in the Havoc$\mab$ work~\cite{Wu2022havoc}.
This is inconsistent with several other large benchmarks~\cite{asprone2022, Metzman2021}, where AFL++ ranks among the top fuzzers.
While not using the persistence mode slows down all fuzzers, we expect state-of-the-art gray-box fuzzers to be affected the most, i.e.,\ they would lose their competitive advantage of speed.
This experiment uses the same targets and setup as the previous section.

\begin{table}
    \caption{Relative degradation of edge coverage not using persistent mode (10 runs).}
    \label{tab:file_mem}
    \centering
    \begin{footnotesize}
    \begin{tabular}{lrrrrrr}
    \toprule
    Target   & AFL               & AFL++             & Havoc$\mab$   & Neuzz         & PreFuzz   & \textbf{Neuzz++} \\
    \midrule          
    harfbuzz    & -8.9\%            & -60.2\%           & -2.9\%        & -3.4\%        & -2.6\%    & -2.9\%       \\
    libjpeg     & -2.8\%            & -55.7\%           & -5.9\%        & -8.9\%        & -7.7\%    & -14.6\%      \\
    sqlite      & 0.1\%            & -57.4\%            & -0.3\%        & -4.9\%        & -7.9\%    & -1.7\%       \\
    woff2       & -34.5\%          & -84.3\%            & -40.4\%       & -43.8\%       & -46.8\%   & -0.6\%       \\
    \bottomrule
    \end{tabular}
    \end{footnotesize}
\end{table}

\Cref{tab:file_mem} presents the performance difference when the persistence mode is not used.
This setup reproduces both the protocol and results from Havoc$\mab$~\cite{Wu2022havoc}, the only paper that compares \ac{NPS}-guided fuzzers against AFL++.
As expected, coverage decreases when passing inputs through files and restarting the program for each test case. 
Most interestingly, AFL++ shows the largest slowdown with a consistent coverage loss over 50\%,
while the AFL-based fuzzers mainly show single-digit percentage degradation.
Consequently, not using the recommended persistence mode can distort the ranking of fuzzers in a benchmark.
In our opinion, this setting does not yield a fair or practically relevant comparison.
Worth mentioning here is that Neuzz++ can compensate the performance loss of its base fuzzer AFL++, obtaining more coverage in absolute values.
As conjectured, results indicate that \ac{NPS}-guided fuzzers suffer less under slow operation than other fuzzers.
Despite that, we are still not able to reproduce the performance of Neuzz and PreFuzz against AFL reported in the original papers.

\begin{takeaway}
    AFL++ is most slowed down when not using the~persistent~mode.
\end{takeaway}

\subsection{Bugs Found}
\label{sec:bugs}

The main goal of fuzzing is to find as many unique bugs as possible.
The default coverage-based crash identification mechanism of AFL and AFL++ tends to overcount unique bugs~\cite{klees2018evaluating}.
To improve this behavior, we apply a more precise stack trace-based deduplication algorithm.
We therefore execute each reported crashing input on the target within \ac{GDB} and retrieve all stack frame addresses when the error occurs. 
This list of addresses then serves as a unique identifier of the triggered bug.
Note that deduplication based on stack traces is ineffective when stack overflow errors occur, because the stack frames are then corrupted.

\begin{table}
    \caption{Bugs found after stack trace deduplication.}
    \label{tab:crash_analysis}
    \centering
    \begin{footnotesize}
    \begin{tabular}{lrrrrrr}
    \toprule
    Target      & AFL   & AFL++ & Havoc$\mab$   & Neuzz & PreFuzz   & \textbf{Neuzz++} \\
    \midrule
    bloaty      & 1     &  1    & \textbf{2}             &  0    & 0         & 1 \\
    guetzli     & 8     & \textbf{264}   & 5             & 0     & 0         & 170 \\
    harfbuzz    & 0     & \textbf{355}   & 1             & 0     & 0         & 12 \\
    json        & 20    & 11    & \textbf{22}            & 18    & 16        & 10 \\
    lcms        & 0     & \textbf{16}    & 0             & 0     & 0         & 8 \\
    libarchive  & 0     & \textbf{1}     & 0             & 0     & 0         & 0 \\
    libxml2     & 0     & \textbf{648}   & 1             & 0     & 0         & 289 \\
    openssl     & 138   & \textbf{1324}  & 409           & 37    & 40        & 721 \\
    pcre2       & 87    & \textbf{4174}  & 262           & 40    & 35        & 1371 \\
    re2         & 0     & \textbf{172}   & 1             & 0     & 0         & 2 \\
    vorbis      & 0     & \textbf{2}     & 1             & 0     & 0         & 0 \\
    woff2       & 20    & 361   & \textbf{671}           & 1     & 1         & 172 \\
    \bottomrule
    \end{tabular}
    \end{footnotesize}
\end{table}

\Cref{tab:crash_analysis} contains the number of crashes with unique stack trace signatures across all trials for each target that reported any crashes.
Neuzz and PreFuzz find the lowest number of crashing inputs (none for most targets), followed by AFL, their base fuzzer; Havoc$\mab$ significantly improves over AFL.
AFL++ is most successful in revealing crashes, with most bugs found and all targets covered.
In summary, all \ac{NPS}-based fuzzers find fewer crashing inputs than the  fuzzer they are based upon.

\begin{takeaway}
    \ac{NPS}-guided fuzzers find fewer bugs than standard fuzzers.
\end{takeaway}

\section{Benchmarking \ac{ML}-based Fuzzers}
\label{sec:guidelines}

Fuzzer evaluation is an open research topic abundently studied in recent works~\cite{klees2018evaluating, Bohme2022, Metzman2021, asprone2022}.
A common guideline is that each fuzzer must be tested on multiple programs, using multiple repetitions to account for randomness.
The recommended number of repetitions revolves around 10--20 trials.
Besides the average performance, indicators of variability (i.e., confidence intervals, statistical tests) are necessary to assess the significance of the results.
The main goal of fuzzers is to find bugs, which suggests that unique bugs found in fixed time should be the evaluation metric.
However, since bugs are rather rare, the performance of fuzzers is often measured in code coverage over time.
This may be justified by observations that more code coverage correlates with more bugs found~\cite{Bohme2022}.
To complement these principles, we propose the following practices when evaluating novel machine learning-based fuzzing methods:

\begin{enumerate}
    \item \textbf{Analyze each new component in the fuzzing loop.}
    Both performance evaluations and ablation studies of \ac{ML} models are critical.
    Metrics specific to the task solved should be used (e.g., accuracy, or precision and recall for classification, mean absolute error or mean squared error for regression, etc.).
    These complement the view on the overall system performance, i.e., coverage or bugs found in the case of fuzzing.
    \ac{ML} evaluation should employ a validation set distinct from the training data to avoid an overly optimistic estimates~\cite{hastie2009}.

    \item \textbf{Use state-of-the-art fuzzers and configurations as baselines.} 
    Lacking strong baselines prevents one from claiming novel state-of-the-art accomplishments in terms of code coverage and bugs found.
    All fuzzers in an experiment should be configured for performance (e.g., appropriate compiler, compilation options, harness, input feeding mode).
    We also recommend introducing new scientific or technical contributions based on recent fuzzers and evaluation platforms, as opposed to their older counterparts.

    \item \textbf{Use comparable metrics for fuzzing performance.}
    As not all fuzzers measure the same type of coverage, we encourage the use of one common evaluation metric between multiple fuzzers.
    In practice, this is easiest done by replaying the corpus at the end of a fuzzing trial, as implemented by FuzzBench~\cite{Metzman2021,asprone2022} and MLFuzz.

    \item \textbf{Repeat trials often enough to account for variance.}
    We propose to use 30 trials for fuzzing evaluation, resulting in tight confidence intervals.
    This sample size is commonly used in statistics and deemed sufficient for the central limit theorem~\cite{chang2006} to hold.
    As shown in \Cref{fig:cov}, \ac{ML}-based fuzzers can have higher coverage variability than gray-box fuzzers, thus requiring more trials for stable baselining.

    \item \textbf{Ensure reproducible results by fixing and serializing parameters.}
    While it is difficult to control all sources of randomness when training \ac{ML} models on GPUs, it remains a good practice in both \acl{ML} and software testing to control possible sources of randomness by seeding random number generators and reusing the same seeds.
    Experimental configurations and, in the case of \ac{ML}, hyperparameters should be documented for reproducibility.

    \item \textbf{Ensure usability of proposed fuzzers.}
    It should be possible to run a newly proposed fuzzer on programs outside the original publication study. 
    Providing a containerized environment can sustainably decrease setup efforts.
    We also support integration of new fuzzers with existing benchmarking platforms, such as FuzzBench and now MLFuzz.
\end{enumerate}

\section{Conclusion and Consequences}
\label{sec:conclusion}
\balance


Neural program smoothing for fuzzing neither reaches its advertised performance, nor does it surpass older fuzzing techniques that are still state-of-the-art.
In our in-depth analysis of \ac{NPS} fuzzers, we
analyzed conceptual limitations of previously published approaches, as well as implementation and evaluation issues.
Our comprehensive benchmark showed that \ac{NPS}-guided fuzzers were by far unable to reach their stated performance.
Addressing the implementation issues did not suffice to outperform state-of-the-art gray-box fuzzers.
The reason for the limited fuzzing performance lies in the difficulty of the \acl{ML} task, which yields trivial models on the data available during fuzzing.

To guide future fuzzing research and practical validation, we developed improved experimental guidelines targeting fuzzing with \acl{ML}.
Our MLFuzz framework for \ac{ML}-based fuzzers includes patched and containerized versions of the investigated fuzzers to help with additional benchmarking.
We encourage researchers to perform ablation studies and provide deeper insights into the components they introduce in fuzzing.

While we highlight fundamental limitations of \acl{NPS}, whether and how much this technique can enhance fuzzing remains an open topic for future research.
We hope that this work contributes to fair and comprehensive evaluations of future fuzzers, be they \ac{ML}-based or not.

\section{Data Availability}
\label{sec:data}
The open-source implementation of Neuzz++ and MLFuzz, the evaluation setup, and raw results are available at
\begin{center}
\url{https://github.com/boschresearch/mlfuzz}\\
\url{https://github.com/boschresearch/neuzzplusplus}.
\end{center}

\section*{Acknowledgements}
Thanks are due to Josselin Feist and the anonymous reviewers for valuable discussions that have led to paper improvements.
This work was supported by the German Federal Ministry of Education and Research (BMBF, project CPSec -- 16KIS1565 and 16KIS1564K).

\begin{acronym}
    \acro{AUC}[AUC]{area-under-the-curve}
    \acro{CFG}[CFG]{control-flow graph}
	\acro{GDB}[GDB]{GNU debugger}
    \acro{ML}[ML]{machine learning}
    \acro{NN}[NN]{neural network}
	\acro{NPS}[NPS]{neural program smoothing}
    \acro{PR}[PR]{precision-recall}
\end{acronym}

\begin{footnotesize}
    \bibliographystyle{ACM-Reference-Format}
    \bibliography{references}
\end{footnotesize}

\balance
\end{document}